\documentclass[preprint,12pt]{elsarticle}

\usepackage[T3,T1]{fontenc}
\DeclareSymbolFont{tipa}{T3}{cmr}{m}{n}
\DeclareMathAccent{\invbreve}{\mathalpha}{tipa}{16}

\usepackage{graphicx}
\usepackage{multicol}
\usepackage{amsmath,amssymb,amsfonts, mathtools}
\usepackage{mathrsfs}
\usepackage{amsthm, dsfont}
\usepackage{rotating}
\usepackage{appendix}
\usepackage[numbers]{natbib}
\usepackage{changepage,afterpage}
\usepackage{etruscan}
\usepackage{tikz, siunitx, newunicodechar}
\usepackage{booktabs,multirow}
\usepackage[utf8]{inputenc}
\usepackage{hyperref}

\newtheorem{theorem}{Theorem}[section]

\newtheorem{proposition}[theorem]{Proposition}
\newtheorem{corollary}[theorem]{Corollary}
\newtheorem{remark}[theorem]{Remark}

\begin{document}
\begin{frontmatter}



\title{Simultaneous symplectic spectral decomposition of positive semidefinite matrices} %% Article title

\author[1,2]{Rudra R.~Kamat} %% Authoor name
\ead{rudra.poet@gmail.com}

\author[3,4]{Hemant K. Mishra} %% Author name
\ead{hemantmishra1124@gmail.com}

\affiliation[1]{organization={Department of Mathematics, Cornell University},%Department and Organization
            city={Ithaca},
            postcode={14850}, 
            state={New York},
            country={USA}}
\affiliation[2]{organization={Department of Mathematics, King's College London},%Department and Organization
            city={London},
            postcode={WC2R 2LS}, 
            country={UK}}
\affiliation[3]{organization={School of Electrical and Computer Engineering, Cornell University},%Department and Organization
            city={Ithaca},
            postcode={14850}, 
            state={New York},
            country={USA}}
\affiliation[4]{organization={Department of Mathematics and Computing, Indian Institute of Technology (Indian School of Mines)},%Department and Organization
            city={Dhanbad},
            postcode={826004}, 
            state={Jharkhand},
            country={India}}
%% Abstract
\begin{abstract}
    We establish necessary and sufficient conditions on simultaneous symplectic spectral decomposition of a family of $2n \times 2n$ real positive semidefinite matrices with symplectic kernels.
    We also provide a precise algebraic condition on a $2n \times 2n$ real positive semidefinite matrix with symplectic kernel for orthosymplectic spectral diagonalization, which generalizes a known result for positive definite matrices.
    
\end{abstract}


% \begin{highlights}
% \end{highlights}

%% Keywords

\end{frontmatter}
%\tableofcontents

%----------%%----------%%----------%%----------%%----------%%----------%%
\section{Introduction}
\label{sec:intro}

    Let $J \coloneqq I_n \otimes \begin{psmallmatrix}
        0 & 1 \\ -1 & 0
    \end{psmallmatrix}$, where $I_n$ is the $n \times n$ identity matrix.
    $\mathds{R}^{2n}$ equipped with the bilinear form $(x, y) \mapsto x^\top J y$ is known as the standard symplectic space.
    A $2n \times 2n$ real matrix $M$ satisfying $M^\top JM=J$ is called a \emph{symplectic matrix}.
    The set of symplectic matrices, denote by $\operatorname{Sp}(2n)$, forms a subgroup of the special linear group of $\mathds{R}^{2n}$ and is closed under transpose \cite{dms}.
    It is known as the \emph{symplectic group}.
    The symplectic group is analogous to the orthogonal group in the sense that a symplectic matrix preserves the bilinear form of the standard symplectic space, just as an orthogonal matrix preserves the Euclidean inner product.
    Unlike the orthogonal group, the symplectic group is non-compact.
    \sloppy A symplectic analog of the classic spectrum theorem, known as Williamson's theorem \cite{williamson1936algebraic}, states that for every $2n \times 2n$ real positive definite matrix $A$ there exists $M \in \operatorname{Sp}(2n)$ such that $M^T A M = D \otimes I_2,$ where $D$ is a positive diagonal matrix.
    The diagonal entries of $D$ are called the \emph{symplectic eigenvalues} of $A$.
    Williamson's theorem is known to be generalized to positive semidefinite matrices by allowing some of the diagonal entries of $D$ to be zero \cite{jm, sonstykel}.
    We say $P$ and $Q$ \emph{symplectically commute} with each other if $PJQ=QJP$.
    
    It is well-known in the classic matrix theory that a set of two or more symmetric matrices can be simultaneously diagonalized by an orthogonal matrix if and only if the matrices commute with each other.
    We provide a symplectic analog of this classic result for positive semidefinite matrices, in which symplectic commutativity plays the role of commutativity in the classic theory of matrices.
    We also report some consequences of the main result which are interesting in their own right.

    The organization of the paper is as follows.
    We begin Section~\ref{sec:symp_normal_form} by recalling some relevant concepts from standard symplectic space, state a known generalization of Williamson's theorem, and establish a proposition that is useful in the proof of the main result.
    In Section~\ref{sec:results} we state and prove the main result, and also report several interesting consequences of the main result.
    In the last Section~\ref{sec:application}, we discuss applications of the main result in two areas, viz., Gaussian quantum information theory and statistical thermodynamics.

%----------------%%----------------%%----------------%%----------------%%----------------%%----------------%
\section{Symplectic spectral decomposition of positive semidefinite matrices}
\label{sec:symp_normal_form}

    We begin by recalling some relevant concepts from the standard symplectic linear space, for which we refer to the first two chapters of \cite{degosson}.
    The {symplectic orthogonal complement} of any subset $\mathscr{W}$ of $\mathds{R}^{2n}$ is defined as
    \begin{align}
        \mathscr{W}^{\perp_{\operatorname{s}}} \coloneqq \big\{u \in \mathds{R}^{2n}: u^\top J w=0 \ \forall w \in \mathscr{W} \big\}.
    \end{align}
    Moreover, $\mathscr{W}^{\perp_{\operatorname{s}}}$ is also a vector subspace.
    In addition, if $\mathscr{W}$ is a vector subspace then it satisfies
    \begin{align}
        \operatorname{dim}\left(\mathscr{W}^{\perp_{\operatorname{s}}}\right)+\operatorname{dim}\left(\mathscr{W}\right)= 2n,
    \end{align}
    and $\left(\mathscr{W}^{\perp_{\operatorname{s}}}\right)^{\perp_{\operatorname{s}}}=\mathscr{W}$.
    A vector subspace $\mathscr{W}$ of $\mathds{R}^{2n}$ is said to be a symplectic subspace if $\mathscr{W} \cap \mathscr{W}^{\perp_{\operatorname{s}}} = \{0\}$. 
    A basis $\{p_1, q_1,\ldots, p_n, q_n\}$ of $\mathds{R}^{2n}$ is called a \emph{symplectic basis} if it satisfies the following conditions: for all $i,j \in \{1,\ldots, n\}$
    \begin{align}
        p_i^\top J p_j= q_i^\top J q_j=0, \quad  p_i^\top J q_j=\delta_{ij},
    \end{align}
    where $(i,j) \mapsto \delta_{ij}$ is the Kronecker delta function.
    There is a one-to-one correspondence between the set of symplectic bases of $\mathds{R}^{2n}$ and the symplectic group, explicitly given by
   \begin{align}
       \{p_1, q_1,\ldots, p_n, q_n\} \mapsto [p_1 \quad q_1 \quad \cdots \quad p_n \quad q_n].
   \end{align}

    For every $2n \times 2n$ real positive semidefinite matrix $A$ with symplectic kernel, there exists $M \in \operatorname{Sp}(2n)$ such that
    \begin{align}\label{eq:symplectic_decomposition}
        M^T A M &= D \otimes I_2,
    \end{align}
    where $D$ is a diagonal matrix with non-negative diagonal entries \cite{jm, sonstykel}.
    The particular case of positive definite matrices is the well-known Williamson's theorem.
    We shall refer to \eqref{eq:symplectic_decomposition} as a \emph{symplectic spectral decomposition} of $A$.
    If, in addition, $M$ is orthogonal as well, we say \eqref{eq:symplectic_decomposition} is an \emph{orthosymplectic} spectral decomposition of $A$.
    Suppose the columns of $M$ in \eqref{eq:symplectic_decomposition} are given by $p_1, q_1,\ldots, p_n, q_n$ and the diagonal entries of $D$ are given by $d_1,\ldots, d_n$.
    Then the relation \eqref{eq:symplectic_decomposition} can be equivalently recast as the following set of equations:
    \begin{align}\label{eq:equivalence_symplectic_decompostion}
        A p_i = d_i J q_i, \qquad A q_i = -d_i J p_i, \qquad 1\leq i \leq n.
    \end{align}
    The proposition given below will be useful in proving the main result.
    {
    \begin{proposition}\label{prop:hamilton_map_diagonalizable_pure_im_eigenvalues}
    Suppose $A$ is a $2n \times 2n$ real positive semidefinite matrix with symplectic kernel.
    Then $JA$ is diagonalizable over $\mathds{C}^{2n}$ and all its eigenvalues are purely imaginary.
    \end{proposition}
    }
    \begin{proof}
        We know that there exists a symplectic basis $\{p_1,q_1,\ldots, p_n,q_n\}$ of $\mathds{R}^{2n}$ and non-negative numbers $d_1,\ldots, d_n$ such that, for all $i \in \{1,\ldots, n\}$,
        \begin{align}
        A p_i &= d_i J q_i, \\
        A q_i &= -d_i J p_i.
    \end{align}
    This implies
        \begin{align}
            JA (p_i + \iota q_i)
                &= -\iota d_i (p_i + \iota q_i), \qquad 1 \leq i \leq n.
        \end{align}
    Here $\iota \coloneqq \sqrt{-1}$.
    So, $p_i \pm \iota q_i$ are eigenvectors of $JA$ corresponding to its  eigenvalues $\mp \iota \mu_i$ for all $1 \leq i \leq n$.
    The fact that $\{p_1,q_1,\ldots, p_n, q_n\}$ is a symplectic basis implies that the set $\{p_i \pm \iota q_i: 1\leq i \leq n\}$ of eigenvectors of $JA$ is linearly independent.
    Therefore, $JA$ is diagonalizable over $\mathds{C}^{2n}$ and all its eigenvalues are purely imaginary.
    \end{proof}

\section{Results}
    \label{sec:results}
    In the following main result, we present precise conditions for simultaneous symplectic spectral decomposition of a family of two or more positive semidefinite matrices with symplectic kernels\footnote{During the preparation of the manuscript, we were made aware of the following online thread of discussions on \emph{mathoverflow} in connection to our main result: \url{https://mathoverflow.net/questions/327421/symplectic-equivalent-of-commuting-matrices}. The discussion reports symplectic commutativity as a condition and requires one of the matrices to be positive definite, to achieve simultaneous symplectic spectral decomposition of two matrices.}.
    We denote the kernel of a matrix $A$ by $\operatorname{ker}(A)$.
    {
    \begin{theorem}\label{thm:simul_symp_normal_form}
    \begin{enumerate}
        \item [(a)] Let $A, B$ be $2n \times 2n$ real positive semidefinite matrices with symplectic kernels. 
        Then $A, B$ can be simultaneously brought to their symplectic spectral decompositions by a common symplectic matrix if and only if $A, B$ symplectically commute with each other and the intersection of the kernels of $A$ and $B$ is also symplectic.
        \item [(b)] Let $\mathcal{F}$ be a non-empty family of $2n \times 2n$ real positive semidefinite  matrices, each of which has symplectic kernel. 
        The matrices in $\mathcal{F}$ can be simultaneously brought to their symplectic spectral decompositions by a common symplectic matrix if and only if the matrices in $\mathcal{F}$ symplectically commute with each other and $\bigcap_{A \in \mathcal{F}} \operatorname{ker}(A)$ is a symplectic subspace.
    \end{enumerate}
    
    \end{theorem}
    }
    \begin{proof}    
        We first prove part $(a)$.
        To prove the only if direction, suppose there exists $M \in \operatorname{Sp}(2n)$ such that
        \begin{align}
            M^\top A M &=  D_1 \otimes I_2, \label{eq:diagonal_form} \\
            M^\top B M &=  D_2 \otimes I_2,
        \end{align}
        where $D_1, D_2$ are $n \times n$ diagonal matrices with non-negative diagonal entries.
        Observe that $J$ commutes with matrices of the form $D \times I_2$, for any $n \times n$ diagonal matrix $D$.
        We thus have
        \begin{align}
            AJB &= M^{-\top} \left( M^\top A M J M^\top B M \right) M^{-1}\\
                &= M^{-\top} \left( D_1 \otimes I_2 \right) J \left(D_2 \otimes I_2 \right) M^{-1}\\
                &= M^{-\top} \left( D_2 \otimes I_2 \right) J \left(D_1 \otimes I_2 \right) M^{-1}\\
                &= M^{-\top} \left( M^\top B M J M^\top A M \right) M^{-1}\\
                &= BJA.
        \end{align}
        Also, suppose the columns of $M$ are $p_1, q_1,\ldots, p_n, q_n$, the diagonal entries of $D_1$ are $\lambda_1,\ldots, \lambda_n$, and the diagonal entries of $D_2$ are $\mu_1,\ldots, \mu_n$. 
        From the equivalent form of the symplectic spectral decomposition given in \eqref{eq:equivalence_symplectic_decompostion}, we thus get
    \begin{align}
        \operatorname{ker}(A) \cap \operatorname{ker}(B) 
            &= \operatorname{span}\{p_i, q_i: i \in \{1,\ldots, n\} \wedge \lambda_i=\mu_i=0\},
    \end{align}
    which is clearly a symplectic subspace.
    
    For the if direction, suppose $AJB=BJA$ and that $\operatorname{ker}(A) \cap \operatorname{ker}(B)$ is a symplectic subspace.
    In what follows, we produce a symplectic matrix $M \in \operatorname{Sp}(2n)$ that diagonalizes both $A$ and $B$ in the sense of \eqref{eq:symplectic_decomposition}.
    \sloppy
    Let $\mathscr{W}\coloneqq \left(\operatorname{ker}(A) \cap \operatorname{ker}(B)\right)^{\perp_{\operatorname{s}}}$, which is a symplectic subspace because $\operatorname{ker}(A) \cap \operatorname{ker}(B)$ is given to be symplectic.
    % Let $A_{\vert \mathscr{W}}$ denote the restriction of $A$ to the symplectic subspace $\mathscr{W}$.
    Moreover, $\mathscr{W}$ is invariant under $JA$ and $JB$, which implies that its complex extension $\mathscr{W}+\iota \mathscr{W}$ is also invariant under these matrices.
    The condition $AJB=BJA$ implies that $JA$ and $JB$ commute with each other.
    Also, it follows by Proposition~\ref{prop:hamilton_map_diagonalizable_pure_im_eigenvalues} that these matrices are diagonalizable over $\mathds{C}^{2n}$.
    Thus, $JA$ and $JB$ are simultaneously diagonalizable over $\mathscr{W}+\iota \mathscr{W}$.
    So, there exists a common eigenvector $u_1+\iota v_1 \in \mathscr{W}+\iota \mathscr{W}$ corresponding some pure-imaginary eigenvalues $\iota \lambda_1$ and $\iota \mu_1$ of $JA$ and $JB$, respectively.
    We then have
    \begin{align}
        JA(u_1+\iota v_1) &= \iota \lambda_1 (u_1+\iota v_1), \label{eq:hamilton_map_eigen_equation_q} \\
         JB(u_1+\iota v_1) &= \iota \mu_1 (u_1+\iota v_1),\label{eq:hamilton_map_eigen_equation_r}
    \end{align}
    which implies
    \begin{align}
        A u_1 &=  \lambda_1 J v_1, \quad A v_1 =  - \lambda_1 J u_1, \label{eq:symp_eigvalue_relations_q} \\
        B u_1 &=  \mu_1 J v_1, \quad B v_1 =  - \mu_1  Ju_1.\label{eq:symp_eigvalue_relations_r}
    \end{align}
    This also gives 
    \begin{align}
        u_1^T A u_1 &= v_1^T A v_1 =  \lambda_1 u_1^T Jv_1, \label{eq:common_symp_eigenvalue_equality_q} \\
        u_1^T B u_1 &= v_1^T B v_1 =  \mu_1 u_1^T Jv_1. \label{eq:common_symp_eigenvalue_equality_r}
    \end{align}
    One of $\lambda_1$ and $\mu_1$ must be non-zero because the intersection of $\mathscr{W}$ and $\mathscr{W}^{\perp_{\operatorname{s}}}$ is trivial.
    Without loss, assume that $\lambda_1 \neq 0$.
    If $\lambda_1 < 0$, then we can rewrite the relations \eqref{eq:symp_eigvalue_relations_q} and \eqref{eq:symp_eigvalue_relations_r} by replacing $\lambda_1$, $\mu_1$, and $u_1$ with their negative values without affecting the further analysis.
    So, there is no loss of generality in assuming that $\lambda_1 >0$.
    Since $A$ is positive semidefinite, the relations in \eqref{eq:common_symp_eigenvalue_equality_q} also imply that $ u_1^T Jv_1 > 0$.
    Also, since $B$ is positive semidefinite, the relations in  \eqref{eq:common_symp_eigenvalue_equality_r} then imply that $\mu_1 \geq 0$.
    We have $\mathscr{W}_1 \coloneqq \operatorname{span}\{u_1, v_1\}$ a symplectic space invariant under both $JA$ and $JB$.
    Choose $p_1 \coloneqq  u_1/\sqrt{u_1^TJ v_1}$ and $q_1 \coloneqq  v_1/\sqrt{u_1^TJ v_1}$ so that $\{p_1, q_1\}$ is a symplectic basis of $\mathscr{W}_1$ and satisfies
    \begin{align}
        A p_1  &= \lambda_1 J q_1, \quad A q_1 =  - \lambda_1 J p_1, \label{eq:symp_eigvalue_relations_q-normalized} \\
        B p_1 &= \mu_1 J q_1, \quad B q_1 =  -\mu_1 J p_1.\label{eq:symp_eigvalue_relations_r-normalized}
    \end{align}

    Let $\mathscr{W}^{\prime} \subset \mathscr{W}$ denote the symplectic orthogonal complement of $\mathscr{W}_1$ in $\mathscr{W}$.
    The relations \eqref{eq:symp_eigvalue_relations_q-normalized} and \eqref{eq:symp_eigvalue_relations_r-normalized} readily imply that $\mathscr{W}^{\prime}$ is invariant under both $JA$ and $JB$.
    We can now repeat the same process for $\mathscr{W}^\prime$, as for $\mathscr{W}$, to get a $2$-dimensional symplectic subspace $\mathscr{W}_2 \subset \mathscr{W}^{\prime}$ with a symplectic basis $\{p_2, q_2\}$ and non-negative numbers $\lambda_2, \mu_2$ satisfying
    \begin{align}
        A p_2 &= \lambda_2 J q_2, \quad A q_2 =  - \lambda_2 J p_2, \\
        B p_2 &= \mu_2 J q_2, \quad B q_2 =  -\mu_2 J p_2.
    \end{align}
    Continue this till $k \coloneqq \frac{1}{2}\operatorname{dim}(\mathscr{W})$ steps to get a symplectic basis $\{p_1, q_1,\ldots, p_k,q_k\}$ of $\mathscr{W}$ and non-negative numbers $\lambda_1,\ldots, \lambda_k$, $\mu_1,\ldots, \mu_k$ satisfying
    \begin{align}
        A p_i &= \lambda_i J q_i, \quad A q_i =  - \lambda_i J p_i, \label{eq:symp_eigvalue_relations_qi-normalized} \\
        B p_i &= \mu_i J q_i, \quad B q_i =  -\mu_i J p_i.
        \label{eq:symp_eigvalue_relations_ri-normalized}
    \end{align}
    for all $i=1,\ldots, k$.
    Let $\{p_{k+1}, q_{k+1}\ldots, p_{n}, q_n\}$ be a symplectic basis of $\mathscr{W}^{\perp_{\operatorname{s}}}= \operatorname{ker}(A) \cap \operatorname{ker}(B)$, and set $\lambda_i\coloneqq 0$, $\mu_i \coloneqq 0$ for $i=k+1,\ldots, n$.
    The relations \eqref{eq:symp_eigvalue_relations_qi-normalized} and \eqref{eq:symp_eigvalue_relations_ri-normalized} mean that the symplectic matrix $M \coloneqq [p_1, q_1, \ldots, p_n, q_n]$ achieves the simultaneous diagonalization of $A$ and $B$ in the sense of \eqref{eq:symplectic_decomposition}.

    Part $(b)$ can be proved by a similar line of arguments and the fact that any non-empty commuting family of diagonalizable linear operators can be diagonalized in a common eigenbasis.
    \end{proof}

\begin{remark}
        In contrast to our work, the paper by Cruz and Faßbender \cite{DELACRUZ2016288} is an interesting read on various conditions for simultaneous diagonalization of matrices via symplectic similarity transformation stated in Theorem~18 of \cite{DELACRUZ2016288}. 
        Our work differs from theirs in the sense that we provide precise conditions for diagonalizability via symplectic \emph{congruence transformation} in the sense of Williamson's theorem.
    \end{remark}

    The following direct consequence of Theorem~\ref{thm:simul_symp_normal_form} is a generalization of a known result on {orthosymplectic spectral decomposition} of positive definite matrices.
    See, e.g., \cite[Proposition~3.7]{son2021symplectic}.
    \begin{corollary}\label{cor:orthosymplectic-diagonalization-williamson}
        Orthosymplectic spectral decomposition of a $2n \times 2n$ real positive semidefinite matrix $A$ with symplectic kernel exists if and only if $JA=AJ$.
    \end{corollary}
    \begin{proof}
        It follows directly by choosing $B$ to be the identity matrix in Theorem~\ref{thm:simul_symp_normal_form}.
    \end{proof}

    The following discussion is for positive definite matrices only.
    We know from the classic matrix theory that if two positive definite matrices commute with each other then their powers also commute.
    Interestingly, it is not the case with symplectic commutativity. 
    Consider the simple example of $A=B=\left(\begin{smallmatrix} 2 & 1 \\ 1 & 1  \end{smallmatrix}\right)$.
    One can verify that $AJB^2 \neq B^2 JA$, even though we have 
    $AJB=BJA$ as well as $AB=BA$.
    This is an instance where \emph{distinct powers} of two positive definite matrices do not symplectically commute with each other, even though the matrices symplectically commute as well as classically commute with each other.
    We also present an example in which \emph{same powers} of $A$ and $B$ do not symplectically commute with each other under the mere assumption of symplectic commutativity of $A$ and $B$.
    Consider
    \begin{align}
        A\coloneqq
        \begin{pmatrix}  
        3&0&0&3  \\ 
        0&8&5&0 \\
        0&5&5&0 \\
        3&0&0&8 
        \end{pmatrix}, \qquad
        B\coloneqq
        \begin{pmatrix}  
        7&0&0&7  \\ 
        0&9&2&0 \\
        0&2&2&0 \\
        7&0&0&9 
        \end{pmatrix}.   
    \end{align}
    One can verify that $AJB=BJA$ but $A^2JB^2 \neq B^2 JA^2$.
    Interestingly, it turns out that the symplectic commutativity of the same powers of $A$ and $B$ can be guaranteed under an additional assumption that $A$ and $B$ classically commute with each other, as stated in the following theorem.

    \begin{theorem}\label{thm:symplectic-commute-powers}
        Let $A, B$ be $2n \times 2n$ real positive definite matrices.
        If $AJB=BJA$ and $AB=BA$, then we have $A^s J B^s= B^s J A^s$ for all $s \in \mathds{R}$.
    \end{theorem}
    \begin{proof}
        The condition $AJB=BJA$ implies that
        \begin{align}\label{eq:symplectic-commutativity-modified}
            B^{-1}AJ=JAB^{-1}.
        \end{align}
        Also, since $A$ and $B$ commute, the matrix $AB^{-1}$ is a symmetric positive definite matrix. 
        Therefore, combining \eqref{eq:symplectic-commutativity-modified} and Corollary~\ref{cor:orthosymplectic-diagonalization-williamson}, we get that $AB^{-1}$ is orthosymplectically diagonalizable in the sense of Williamson's theorem.
        This implies that for any $s \in \mathds{R}$ the matrix $A^sB^{-s}$ is also orthosymplectically diagonalizable in the sense of Williamson's theorem.
        By invoking Corollary~\ref{cor:orthosymplectic-diagonalization-williamson} again, we thus have $JA^sB^{-s}=A^sB^{-s}J$.
        Using the commutativity of $A$ and $B$, this simplifies to $A^s J B^s = B^s J A^s$.
    \end{proof}

%------------------%%------------------%%------------------%%------------------%
\section{Applications}
\label{sec:application}

    We discuss two applications of the main result in this section. 
    The first application is a characterization of two mean zero Gaussian states to be decomposed into normal modes by a common Gaussian unitary operation.
    The second application is deriving an analytical expression for the partition function in statistical thermodynamics in terms of the symplectic eigenvalues of the positive definite matrix of the associated quadratic Hamiltonian.

\subsection{Normal mode decomposition of Gaussian states by common Gaussian unitary operation}
    An $n$-mode Gaussian quantum state $\rho$ is uniquely determined by its mean vector $r \in \mathds{R}^{2n}$, and its covariance matrix $V$ which is a $2n \times 2n$ real symmetric positive definite matrix \cite{serafini2017quantum}.
    Let $S$ be a symplectic matrix diagonalizing $V$ in the sense of Williamson's theorem.
    Associated with $r$ and $M$ are unitary transoformations on the system of the Gaussian state, known as Weyl displacement operator $\hat{D}_r$ and Gaussian unitary operation $\hat{S}$, respectively. 
    The Gaussian state $\rho$ can be decoupled into a tensor product of thermal states using these unitary transformations:
    \begin{align}\label{eq:normal_mode_dec}
        \hat{D}_{r} \hat{S}^{\dagger} \rho \hat{S} \hat{D}_{-r} = \bigotimes_{i=1}^n \tau_i,
    \end{align}
    where $\tau_i$ are some thermal states \cite[Eq.~3.38]{serafini2017quantum}.
    The relationship \eqref{eq:normal_mode_dec} is known as a normal mode decomposition of the Gaussian state.

    Let $\rho_1$ and $\rho_2$ be mean zero Gaussian states with covariance matrices $V_1$ and $V_2$, respectively.
    We know that a Gaussian unitary can bring $\rho_1$ and $\rho_2$ into their normal mode decomposition forms if and only if $V_1$ and $V_2$ can be simultaneously brought to their symplectic spectral decompositions by a common symplectic transformation.
    By Theorem~\ref{thm:simul_symp_normal_form}, this is equivalent to the algebraic condition $V_1 J V_2 = V_2 J V_1$ on the covariance matrices of the Gaussian state.

\subsection{Analytical expression for the partition function}

   In statistical mechanics, the \textit{partition function} $Z$ is a fundamental quantity used to describe the statistical properties of a system in thermodynamic equilibrium\footnote{The notation $Z$ comes from the German word \emph{Zustandssumme}, which means ``sum of states''.}. 
   The partition function serves as a bridge between the microscopic states of a system and its macroscopic properties, and it is also used to derive various other thermodynamic quantities such as the free energy, entropy, internal energy, and specific heat of the system.
    The partition function of a gas of $N$ identical classical particles in $d$ dimensions is given by \cite[Chapter~7]{Huang1967}:
    \begin{equation}
        Z=\frac{1}{N! \ h^{dN}} \int_{\mathds{R}^{2dN}}\! d^{2dN}(p,q) \ \ \exp\left[-\beta H(p, q) \right],
    \end{equation}
    where $(p,q) \in \mathds{R}^{2dN}$ indicate the momenta and positions of the particles in a $d$-dimensional space, respectively; $h$ is Planck's constant, and $H$ is the Hamiltonian of the system.
    We consider the case where the Hamiltonian $H$ is quadratic and positive definite. Such a Hamiltonian is generally given by \cite[Eq.~1]{Härkönen_2022}
    and the partition function then takes the form
    \begin{align}\label{eq:Z_integral_form}
        Z=\frac{1}{N!\ h^{dN}} \int_{\mathds{R}^{2Nd}}\! d^{2dN}(z) \ \   \exp\left[ - \frac{\beta}{2}z^{T} \left(\sum_{{i=1}}^{{N}} M_{i} \right)z \right],
    \end{align}
    where $M_1,\ldots, M_N$ are $2dN \times 2dN$ real positive definite matrices.
    Under the condition that $M_1,\ldots, M_N$ pairwise symplectically commute with each other, we get using Theorem~\ref{thm:simul_symp_normal_form} a symplectic matrix $S$ diagonalizing $M_i$ in the sense of Williamson's theorem as $S^T M_i S= \operatorname{diag}(d^{[i]}_{1},\ldots, d^{[i]}_{dN}) \otimes I_2$ for $1 \leq i \leq N$.
    By substituting these decompositions in \eqref{eq:Z_integral_form} and then applying the  Gaussian integral formula, we get the following analytical expression of $Z$ in terms of the symplectic eigenvalues of $M_1,\ldots, M_N$:
    \begin{align}
        Z 
        = \left(\frac{\pi}{\beta h} \right)^{dN} \left(N! \  \prod_{j=1}^{dN}  \left[ \sum_{i=1}^{N} d^{[i]}_j \right] \right)^{-1}.
    \end{align}

\section{Future directions}
\label{sec:future}
    An interesting future work would be to prove an analog of Theorem~\ref{thm:simul_symp_normal_form} for Williamson's theorem in the infinite-dimensional case, which was developed in \cite{bhat2019real}.
    Another potential application of the main result is in physical systems with quadratic integrals of motion, such as those with identical particles classified in \cite{Brihaye2004}. The phase space trajectories in such systems are constrained to lie on the level surfaces of the integrals of motion. Our main result can be applicable when the physical system in question is also integrable, i.e., when the quadratic integrals of motion form a Poisson commuting family.
    In such cases, it is easier to analyze the stability of the physical system because the integrals of motion can be reduced to a normal form in a common symplectic basis.

\section*{Declaration of competing interest}
The author declares that there is no competing interest.

\section*{Data availability}
No data was used for the research described in this article.

\section*{Declaration of generative AI and AI-assisted technologies in the writing process}
No AI or AI-assisted technologies were used in the writing process.

\section*{Acknowledgements}
    HKM acknowledges supports from the NSF under grant number 2304816, AFRL under agreement number FA8750-23-2-00, and FRS Project No.~MISC $0147$.
    RRK would like to thank Prof.~Maurice de Gosson foremost for suggesting a closely related summer project topic that culminated into this work.
    RRK also thanks Prof.~Reyer Sjamaar for his supervision on an undergraduate thesis of the author which reviewed the literature on Williamson's theorem.

\end{document}